\documentclass{article}

\usepackage{amsmath}
\usepackage{amsthm}
\usepackage{graphicx}
\usepackage{amssymb}
\usepackage{psfrag}

\newcommand{\tr}{\operatorname{tr}}
\newcommand{\conv}{\operatorname{conv}}
\newcommand{\conc}{\operatorname{conc}}

\newcommand{\gr}{\operatorname{gr}}

\newcommand{\rank}{\operatorname{rank}}
\newcommand{\uinvnorm}{|\kern-2pt|\kern-2pt|}

\theoremstyle{plain}
\newtheorem{theorem}{Theorem}[section]
\newtheorem{lemma}[theorem]{Lemma}
\newtheorem{proposition}[theorem]{Proposition}
\newtheorem{corollary}[theorem]{Corollary}

\theoremstyle{definition}
\newtheorem{definition}[theorem]{Definition}

\theoremstyle{remark}
\newtheorem{remark}[theorem]{Remark}

\begin{document}

\bibliographystyle{amsalphaxxx}

\title{{\sf\bfseries Convex Hulls of Varieties and Entanglement Measures Based on the Roof Construction}}
\author{{\sf Tobias J.\ Osborne}\footnote{\texttt{T.J.Osborne@bristol.ac.uk}}\\ {\sf Centre for Quantum Computer Technology} \\ {\sf and Department of Physics}\\ {\sf The University of Queensland, 4072}\\ {\sf Australia}\\ {\sf and} \\ {\sf School of Mathematics}\\ {\sf University of Bristol}\\ {\sf University Walk, Bristol BS8 1TW}\\ {\sf United
Kingdom}}
\date{\today}
\maketitle

\begin{abstract}
In this paper we study the problem of calculating the convex hull
of certain affine algebraic varieties. As we explain, the
motivation for considering this problem is that certain pure-state
measures of quantum entanglement, which we call \emph{polynomial
entanglement measures}, can be represented as affine algebraic
varieties. We consider the evaluation of certain mixed-state
extensions of these polynomial entanglement measures, namely
\emph{convex and concave roofs}. We show that the evaluation of a
roof-based mixed-state extension is equivalent to calculating a
hyperplane which is multiply tangent to the variety in a number of
places equal to the number of terms in an optimal decomposition
for the measure. In this way we provide an \emph{implicit}
representation of optimal decompositions for mixed-state
entanglement measures based on the roof construction.
\end{abstract}

\section{Introduction}
The burgeoning field of quantum information science is beginning
to provide us with hints that quantum systems are capable of
performing spectacularly powerful information processing tasks. An
example of this is the existence of a \emph{quantum algorithm},
Shor's algorithm, which can factor in polynomial time
\cite{shor:1994a, shor:1997a}. A consequence of discoveries such
as Shor's factoring algorithm is the emergence of a wide\-spread
belief that, parallel to the physical resource of
\emph{information}, there is a corresponding resource which
quantum systems support called \emph{quantum information}. The
study of how quantum information can be manipulated and processed
is the central goal of quantum information science.

So what is quantum information? It is folklore within the quantum
information community that it may be \emph{quantum entanglement},
the unique property possessed by quantum states of bipartite
systems not expressible as simple tensor products. A classic
example of an entangled state is the singlet $|\Psi^-\rangle =
\frac{1}{\sqrt2}(|01\rangle - |10\rangle)$, which is commonly
considered to be the \emph{fundamental unit} of entanglement.
There are a number of fascinating information processing tasks
which use entangled states as a resource, such as teleportation
\cite{bennett:1993a} and enhanced classical communication
\cite{nielsen:2000a}.

Throughout the progress of the research conducted in quantum
information science, in particular in the study of quantum
entanglement, a number of fascinating mathematical problems have
been encountered. Within the theory of quantum entanglement and
quantum communication these problems can be broadly described as
the the classification of positive operators on tensor products of
vector spaces. This classification problem is central to these
theories. Many of the stumbling blocks currently encountered by
researchers in quantum information science are due to incomplete
results in this classification program.

Despite the many problems encountered, a fruitful interplay
between mathematics and physics has already resulted from progress
in this classification program. An example of this is the recent
work on indecomposable positive linear maps. In this case a result
obtained previously by Woronowicz \cite{woronowicz:1976a} was
applied by Pawe{\l} Horodecki to construct quantum states which
have the property that they are \emph{bound entangled}
\cite{horodecki:1997a} (for a review of bound entanglement and
distillable entanglement see the paper \cite{horodecki:2001b}).
The subsequent work on bound entanglement was then fed back into
mathematics by Terhal \cite{terhal:2001a}, who used this work in
physics to construct new indecomposable positive linear maps.

Bound entanglement is a particular nonlocal phenomena exhibited by
certain mixed states\footnote{Mixed states are probabilistic
mixtures of quantum states.} which falls within the framework of
mixed-state entanglement, or the study of nonlocal quantum
features of mixed states. Understanding mixed-state entanglement
is an immediate and pressing priority in quantum information
science. This is because a number of expected developments in
areas such as quantum communication cannot take place until
certain questions about mixed-state entanglement are answered.

Quite apart from the ramifications for quantum information, we
believe that the study of mixed-state entanglement will lead to
new and interesting results in mathematics. In particular, there
are suggestive indications (see, for example, the work on
\emph{concurrence} for some striking results \cite{hill:1997a,
wootters:1998a, uhlmann:2000a, audenaert:2001a, osborne:2002b})
that there are deep results waiting to be found in the field of
matrix analysis.

The purpose of this paper is to provide a method to evaluate
certain functions on the space of positive operators, called
mixed-state entanglement measures. To do so, in
\S\ref{sec:prelims} we begin by establishing some results in
convexity theory. These results allow us to state our main result,
a necessary condition for constructing convex hulls of certain
subsets. In \S\ref{sec:poincare} we make the observation that
quantum state-space can be represented by the convex hull of a
quadratic algebraic variety. Finally, in \S\ref{sec:entmeas} we
combine these two results to provide an implicit representation
for a family of mixed-state entanglement measures, namely those
based on the roof construction.

\section{Preliminaries}\label{sec:prelims}
In this section we introduce the main object of study for this
paper: the convex or concave roof. We also establish some general
results concerning the construction and evaluation of roofs, the
most important result being Proposition~\ref{thm:spanub}. We
illustrate the definitions and results throughout this section in
terms of a simplified example.

\subsection{Properties of Convex and Concave Roofs}
We begin by establishing some notation. Let $V$ denote the subset
of $\mathbb{R}^n$ formed from the intersection of the zero-sets of
a collection of polynomials $\ell_0,\ldots,\ell_a$ $\in$ $
\mathbb{R}[x_1,\ldots,x_n]$, which we write as
\begin{equation}
V=\mathcal{Z}(\ell_0,\ldots,\ell_a).
\end{equation}
The set $V$ is known as an \emph{affine algebraic variety}. We
assume that this set is compact in the standard topology on
$\mathbb{R}^n$, irreducible, and nonsingular, i.e., $V$ is a
nonsingular variety. (For a gentle introduction to varieties and
algebraic geometry, including definitions of the terms we use
here, see \cite{cox:1997a} and \cite{harris:1995a}. For a more
complete reference see \cite{hartshorne:1997a}.) We have assumed
that $V$ is irreducible and nonsingular in order to avoid various
pathologies associated with the definition of the tangent space
for these spaces. It should be noted that it is possible, with a
little extra work, to extend the results of this section to
include these cases. However, for our applications the assumption
of nonsingularity imposes no limitations, so we ignore singular
varieties.

\begin{figure*}
\begin{center}
\psfrag{-1}{$-1$} %
\psfrag{0}{$0$} %
\psfrag{1}{$1$} %
\psfrag{f}{$f$} %
\psfrag{x}{$x$} %
\psfrag{y}{$y$} %
\includegraphics{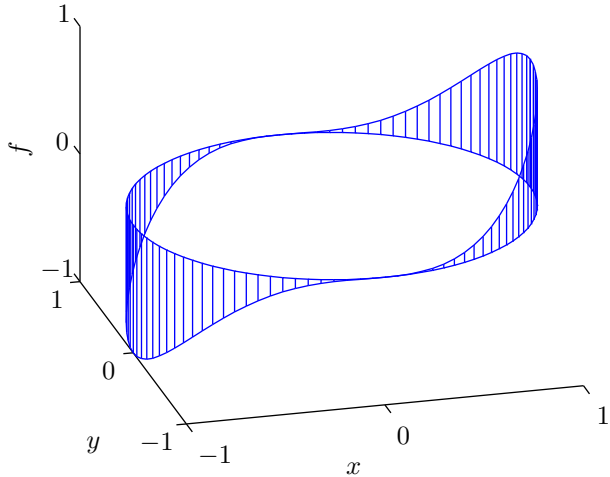}
\caption{The graph $(\gr f)(\mathbf{r})$ of $f(x,y)=x^3$ on
$\mathcal{Z}(x^2+y^2-1)$. The graph of $f$ is a space curve in
three dimensions. We have indicated this by placing vertical lines
connecting the set $\mathcal{Z}(x^2+y^2-1)$ with the graph. This
representation is adopted to reinforce the fact that the graph of
$f$ is plotted ``above''
$\mathcal{Z}(x^2+y^2-1)$.}\label{fig:frame}
\end{center}
\end{figure*}

Let $f$ be an arbitrary continuous scalar-valued function defined
on $V$. (The assumption of continuity is invoked so that
subsequent constructions are at least continuous.) It suffices for
most of our applications to take this function to be a polynomial,
in which case $f$ is an element of the coordinate ring of $V$,
$f\in\mathbb{R}[V]$. An example variety $V$ and function $f$ is
shown in Figure~\ref{fig:frame}. The objective of this paper is to
construct and evaluate a scalar-valued function $g$ which is equal
to $f$ on $V$, but which is defined over all of $\conv V$, the
convex hull of $V$. Obviously there are many ways to define such
functions. To proceed, we single out two special possibilities for
$g$, defined by the formulas:
\begin{equation}\label{eq:caraconvroof}
(\conv f)(\mathbf{r}) \triangleq \inf\Biggl\{ \sum_{j=1}^{m} p_j
f(\mathbf{x}_j)\,\Bigg\lvert\,
 \mathbf{x}_j \in V,  \sum_{j=1}^{m} p_j\mathbf{x}_j=\mathbf{r}
 \Biggr\}
\end{equation}
and
\begin{equation}\label{eq:caraconcroof}
(\conc f)(\mathbf{r}) \triangleq \sup\Biggl\{ \sum_{j=1}^{m} p_j
f(\mathbf{x}_j)\,\Bigg\lvert\,
 \mathbf{x}_j \in V,  \sum_{j=1}^{m} p_j\mathbf{x}_j=\mathbf{r}
 \Biggr\},
\end{equation}
where $\mathbf{r}$ is an arbitrary point in $\conv V$,
$p_j\in\mathbb{R}$ is a probability distribution, and $m$ is an
arbitrary positive integer. A simple consequence of
Carath\'eodory's Theorem is that $m$ is bounded above by the
dimension of $V$, $m\le v=\dim V$ \cite{rockafellar:1970a}.
\begin{definition}[\cite{uhlmann:1998a}] The
functions $\conv f$ and $\conc f$ defined by
(\ref{eq:caraconvroof}) and (\ref{eq:caraconcroof}) are called the
\emph{convex roof} and \emph{concave roof} of $f$, respectively.
We refer to the procedure of constructing $\conv f$ or $\conc f$
as \emph{roofing} $f$.
\end{definition}

One of our main objectives is to evaluate $\conv f$ or $\conc f$
at an arbitrary point $\mathbf{r}\in\conv V$. Obviously this can
be done by simply carrying out the optimisation in
(\ref{eq:caraconvroof}) and (\ref{eq:caraconcroof}).
Unfortunately, evaluating $\conv f$ at a point $\mathbf{r}\in
\conv V$ according to (\ref{eq:caraconvroof}) and
(\ref{eq:caraconcroof}) is, in general, very difficult. For this
reason, in this section, we provide an alternative way to evaluate
these functions.

\begin{figure*}
\begin{center}
\psfrag{-1}{$-1$} %
\psfrag{-0.5}{$-0.5$} %
\psfrag{0}{$0$} %
\psfrag{0.5}{$0.5$} %
\psfrag{1}{$1$} %
\psfrag{convf}{$\conv(\gr f)$} %
\psfrag{x}{$x$} %
\psfrag{y}{$y$} %
\includegraphics{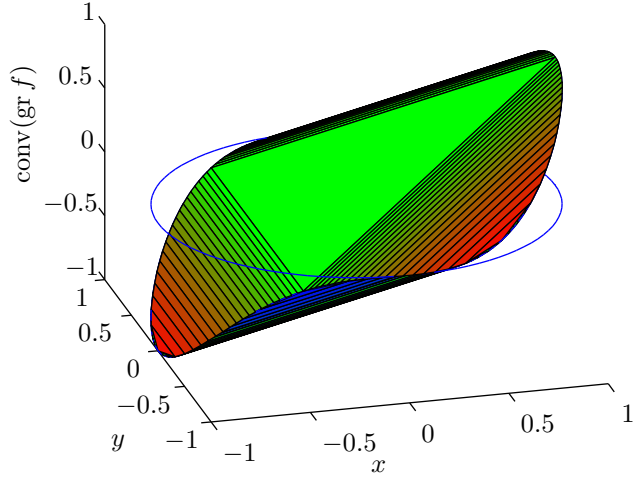}
\caption{The three dimensional convex hull $\conv(\gr f)$ of the
graph of $f(x,y)=x^3$ on $\mathcal{Z}(x^2+y^2-1)$. Note that this
is a three-dimensional convex set. The upper boundary is shaded
red through green and the lower boundary is shaded
blue.}\label{fig:convhull}
\end{center}
\end{figure*}

We shall spend a lot of time discussing properties of the
\emph{graph} of $f$ on $V$, denoted $\gr f$, which is the subset
of $\mathbb{R}^{n+1}$ defined by
\begin{equation}
\gr f \triangleq \{(\mathbf{r},f(\mathbf{r})) \,|\, \mathbf{r}\in
V \}.
\end{equation}
Associated with $\gr f$ is the convex hull of the graph of $f$,
which is the subset of $\mathbb{R}^{n+1}$ given by
\begin{equation}\label{eq:convepif}
\conv(\gr f) \triangleq \Biggl\{ (\mathbf{x},\mu) \,\Bigg|\,
(\mathbf{x},\mu)=\sum_{j=1}^{v'+1}p_j
(\mathbf{x}_j,f(\mathbf{x}_j))\Biggr\},
\end{equation}
where $\mathbf{x}_j\in\mathbb{R}^n$, $p_j$ is a probability
distribution, and we have invoked Carath\'eodory's Theorem to
bound the number of terms in the sum by $v'+1=\dim(\conv V)+1$.
See Figure~\ref{fig:convhull} for an example of what $\conv(\gr
f)$ looks like for the variety $V$ and function $f$ introduced in
Figure~\ref{fig:frame}.

\begin{figure*}
\begin{center}
\psfrag{-1}{$-1$} %
\psfrag{-0.5}{$-0.5$} %
\psfrag{0}{$0$} %
\psfrag{0.5}{$0.5$} %
\psfrag{1}{$1$} %
\psfrag{convf}{$\conv f$} %
\psfrag{x}{$x$} %
\psfrag{y}{$y$} %
\includegraphics{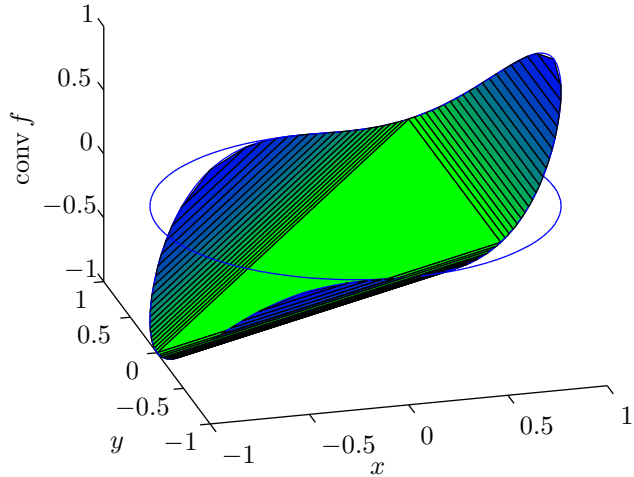}
\caption{Convex roof for the function $f(x,y)=x^3$ on
$\mathcal{Z}(x^2+y^2-1)$. Note that the graph of $\conv f$ is the
lower boundary of the convex set $\conv(\gr f)$.}\label{fig:roof}
\end{center}
\end{figure*}

The connection between $\conv(\gr f)$ and $\conv f$ and $\conc f$
is provided by the following lemma.
\begin{lemma}\label{cor:roofhull}
The convex and concave roofs of $f$ are given by
\begin{equation}\label{eq:altconvroof}
(\conv f)(\mathbf{r}) = \inf\{\mu \,|\, (\mathbf{r},\mu)\in
\conv(\gr f) \}
\end{equation}
and
\begin{equation}\label{eq:altconcroof}
(\conc f)(\mathbf{r}) = \sup\{\mu \,|\, (\mathbf{r},\mu)\in
\conv(\gr f) \},
\end{equation}
respectively.
\end{lemma}
\begin{proof}
The equality of (\ref{eq:altconvroof}) and (\ref{eq:caraconvroof})
(respectively, (\ref{eq:altconcroof}) and (\ref{eq:caraconvroof}))
follows immediately from the definition of (\ref{eq:convepif}).
\end{proof}
\begin{remark}
Lemma~\ref{cor:roofhull} shows us that the convex roof of $f$ is
the ``lower boundary'' of the convex hull of the graph of $f$.
Similarly, the concave roof $\conc f$ is the ``upper boundary'' of
$\conv(\gr f)$. This is illustrated for $\conv f$ in
Figure~\ref{fig:roof}.
\end{remark}

The convex and concave roofs of $f$ are distinguished from all
other functions $g$ on $\conv V$ equal to $f$ on $V$ by the
following result.
\begin{lemma}[Lemma A-3, \cite{uhlmann:1998a}; \cite{uhlmann:2003a}\footnote{Armin Uhlmann has proved (private communication) this lemma for all compact $V$ and continuous $f$.}]\label{lem:uhlmanna3}
The convex roof (respectively, concave roof) of $f$ is, pointwise,
the largest convex (respectively, smallest concave) function equal
to $f$ on $V$.
\end{lemma}

We now turn to the discussion of the main result of this section:
an alternative way to evaluate $\conv f$ and $\conc f$. This
result takes the form of a necessary condition which limits the
space over which the infimum and supremum in
(\ref{eq:caraconvroof}) and (\ref{eq:caraconcroof}) needs to be
taken.

To begin, we make use of the fundamental duality between convex
sets and intersections of halfspaces to write $\conv(M)$, where
$M\subset\mathbb{R}^n$ is some bounded set, as an intersection in
$\mathbb{R}^{n}$ of a collection of halfspaces,
\begin{equation}
\conv(M) = \bigcap_{\mathcal{H}_{(\mathbf{m},\omega)}\in \Omega}
\mathcal{H}_{(\mathbf{m},\omega)},
\end{equation}
where
$\mathcal{H}_{(\mathbf{m},\omega)}=\bigl\{\mathbf{x}\in\mathbb{R}^{n}
\,\big|\, \mathbf{m}\cdot\mathbf{x} \le \omega \bigr\}$ is a
halfspace, and $\Omega$ is the set of all halfspaces containing
$M$. Where obvious we abuse notation and write
$H_{(\mathbf{m},\omega)}\in\Omega$ to refer to the hyperplane
$H_{(\mathbf{m},\omega)}=\bigl\{\mathbf{x}\in\mathbb{R}^{n}
\,\big|\, \mathbf{m}\cdot\mathbf{x} = \omega \bigr\}$ associated
with a halfspace $\mathcal{H}_{(\mathbf{m},\omega)}\in\Omega$.

\begin{definition}
A hyperplane $H$ in $\mathbb{R}^n$ is said to be
$m$-\emph{tangent} to a nonsingular variety (or indeed any
sufficiently well-behaved manifold) $M\subset \mathbb{R}^n$ at
points $\mathbf{x}_1, \ldots, \mathbf{x}_m\in M$ if $H$ contains
both the points $\mathbf{x}_1, \ldots, \mathbf{x}_m\in M$ and
their associated tangent spaces $T_{\mathbf{x}_j}M$. If $m=2$ or
$3$ then $H$ is said to be bitangent or tritangent to $M$,
respectively.
\end{definition}

The set $\Omega$ contains three types of halfspace. Firstly, there
are those halfspaces which completely contain $M$ and whose
associated hyperplane does not contain any point of $M$. Secondly,
there are those halfspaces whose associated hyperplane contains
one or more points of $M$. These halfspaces are said to
\emph{support} $M$. Finally, of those halfspaces which support $M$
at some collection of points $\mathbf{x}_1,\ldots,\mathbf{x}_m$
there are those whose associated hyperplane is $p$-tangent to $M$
at some subset of $p$ points of
$\{\mathbf{x}_1,\ldots,\mathbf{x}_m\}$ (because of the assumption
of nonsingularity, $M$ has a well-defined tangent space at each
point).

One of our main results, for this subsection, is a necessary
condition for the hyperplane $H$ associated with some halfspace
$\mathcal{H}\in\Omega$ to support $M$:
\begin{proposition}\label{thm:spanub}
Let $H\in\Omega$ be a hyperplane that supports a nonsingular
variety (or manifold) $M\subset\mathbb{R}^n$ at $m$ points
$\mathbf{x}_1,\ldots,\mathbf{x}_m\in M$. The (translate of an)
$(n-1)$-dimensional vector space defined by $H$ must contain, as
subspaces, the vector spaces $T_{\mathbf{x}_j}M$, $j=1\ldots,m$
and the vector space formed from the vectors
$\mathbf{x}_j-\mathbf{x}_k$, $ \forall j\not=k=1,\ldots,m$. In
particular, if $H\in\Omega$ supports $M$ at
$\mathbf{x}_1,\ldots,\mathbf{x}_m$, it must be $m$-tangent to $M$
at $\mathbf{x}_1,\ldots,\mathbf{x}_m$.
\end{proposition}
\begin{proof}
A hyperplane $H\in\Omega$ which supports $M$ at $m$ points
$\mathbf{x}_1,\ldots,\mathbf{x}_m\in M$ contains these points so,
in particular, it must contain all the vectors
$\mathbf{x}_j-\mathbf{x}_k$, $\forall j\not=k=1,\ldots,m$. In
addition, if $H$ did not contain $T_{\mathbf{x}_j}M$ as a
subspace, for all $j$, then some part of $H$ would lie inside the
convex hull of $M$ near some point $\mathbf{x}_j$, contradicting
$H\in\Omega$. The second statement is simply a restatement of the
first.
\end{proof}

\begin{remark}
Proposition~\ref{thm:spanub} provides a necessary condition for a
hyperplane $H\in\Omega$ to support $M$. In general this condition
is not sufficient as there exist hyperplanes $H'\not\in\Omega$
which satisfy the condition of Proposition~\ref{thm:spanub}. The
effect of this proposition is to identify a subset of
$\Phi\subset\Omega$ such that the convex hull of $M$ can still be
found by intersecting members of $\Phi$.
\end{remark}

We now specialise the results we have just obtained to describe
the structure of $\conv(\gr f)$. As demonstrated by
Corollary~\ref{cor:roofhull}, the convex and concave roofs of a
function $f$ are the top and bottom halves of the boundary of
$\conv(\gr f)$. It is useful to obtain a characterisation of this
boundary. To do so, we make use of the observations above and a
lemma of Uhlmann \cite{uhlmann:1998a} to describe the nature of
these two hypersurfaces.

It turns out that the surfaces in $\mathbb{R}^{n+1}$ defined by
either $\conv f$ or $\conc f$ are comprised of a union of a family
of polytopes (convex hulls of finite sets of points) which,
pairwise, have zero intersection. To see this\footnote{From now on
we will phrase most of our results in terms of the convex roof of
$f$. This does not entail any loss of generality as all statements
about $\conv f$ extend in an obvious way to $\conc f$.}, note that
any point $(\mathbf{r},\mu)$ of $\conv f$ may be written as a
finite sum $(\mathbf{r},\mu) = \sum_{j=1}^{v+1}
\lambda_j(\mathbf{x}_j,f(\mathbf{x}_j))$. We call this sum an
\emph{optimal $(v+1)$-decomposition} of $\mathbf{r}$ for $\conv
f$. (Where the convexity or concavity is obvious we simply refer
to this sum as an optimal decomposition.) A lemma of Uhlmann
(lemma 3, \cite{uhlmann:1998a}) shows us that something stronger
actually applies, namely that the polytope defined by the convex
hull of the points
$(\mathbf{x}_1,f(\mathbf{x}_1)),\ldots,(\mathbf{x}_{v+1},f(\mathbf{x}_{v+1}))$
also lies entirely within the surface defined by $\conv f$. The
consequence of this observation is that the surface defined by
$\conv f$ (which is one half of the boundary of $\conv(\gr f)$) is
made completely from a (pairwise disjoint) union of polytopes.
Further, each of these polytopes $P$ is a piece of the hyperplane
which supports $\gr f$ at the extreme points
$(\mathbf{x}_1,f(\mathbf{x}_1)),\ldots,(\mathbf{x}_{v+1},f(\mathbf{x}_{v+1}))$
(because the polytope lives on the boundary of the convex set
$\conv(\gr f)$). The structure of $\conv f$ is illustrated for the
example of $f=x^3$ in Figure~\ref{fig:roof}, where, in this case,
it may be seen that $\conv f$ is comprised of line segments and
one triangle.

For a given $\mathbf{r}\in\conv V$, we are interested in
determining the polytope $P$ on the boundary of $\conv(\gr f)$
which contains $(\mathbf{r},f(\mathbf{r}))$. Equivalently, given
$\mathbf{r}\in\conv V$ we are interested in determining an optimal
$m$-decomposition for $\conv f$ at $\mathbf{r}$. One way of
approaching this problem is to simply search the entire space of
all sequences of points
$(\mathbf{x}_1,f(\mathbf{x}_1)),\ldots,(\mathbf{x}_{v+1},f(\mathbf{x}_{v+1}))$
which achieve the infimum in (\ref{eq:caraconvroof}) and check to
see if their convex hull contains $(\mathbf{r},f(\mathbf{r}))$.
Proposition~\ref{thm:spanub} indicates that this is not necessary
because we need only search through sequences of points
$(\mathbf{x}_1,f(\mathbf{x}_1)),$ $\ldots,$
$(\mathbf{x}_{m+1},f(\mathbf{x}_{m+1}))$ for which there exists an
$m$-tangent hyperplane passing through them. This statement is the
content of the following corollary.
\begin{corollary}\label{cor:mtang}
Let $\mathbf{r}\in\conv(V)$. If $\{\lambda_j,\mathbf{x}_j\}$ is an
optimal $m$-decomposition for $\conv f$ at $\mathbf{r}$ then there
is a hyperplane $H$ passing through $\mathbf{x}_j, \forall j$
which is $m$-tangent at $\mathbf{x}_j, \forall j$.
\end{corollary}

\subsection{Calculating Optimal Decompositions for Polynomial Roofs}
In this subsection we utilise Corollary~\ref{cor:mtang} to provide
an implicit representation for an optimal decomposition of a point
$\mathbf{r}\in\conv V$ for $\conv f$. We show that when $f$ is a
polynomial an optimal decomposition for $\conv f$ at a point
$\mathbf{r}\in\conv V$ may be found by simultaneously solving a
set of polynomial equations. In principle this enables us to
invoke the machinery of algebraic geometry, in particular
Gr{\"{o}}bner bases methods, to solve the problem \cite{cox:1997a,
cox:1998a, harris:1995a, hartshorne:1997a}.

Corollary~\ref{cor:mtang} shows that to find an optimal
decomposition for $\conv f$ at a point $\mathbf{r}\in\conv V$ we
need only search over points $\mathbf{x}_1,\ldots,\mathbf{x}_m$
for which there exists an $m$-tangent hyperplane $H$ passing
through these points. If we were able to obtain an expression for
the basis for the tangent space $T_{\mathbf{x}_j}V$ at
$\mathbf{x}_j$ then we could reduce this problem to establishing
the linear dependence of a collection of vectors in the following
way.

Consider the tangent space $T_{\mathbf{x}_j}V$ at a point
$\mathbf{x}_j\in V$. Suppose we had a formula, in terms of the
components of $\mathbf{x}_j$, for each the basis vectors
$\alpha_1^{(j)},\ldots,\alpha_t^{(j)}$ of $T_{\mathbf{x}_j}V$,
where $t=\dim T_{\mathbf{x}_j} V$. Using this formula we could
rephrase Proposition~\ref{thm:spanub} as the statement that for an
$m$-tangent hyperplane $H$ supporting $V$ at
$\mathbf{x}_1,\ldots,\mathbf{x}_m$, the matrix $R$ whose rows are
formed from the vectors $\alpha_1^{(j)},\ldots,\alpha_t^{(j)}$ and
$\mathbf{x}_j-\mathbf{x}_k$, $\forall j\not=k=1,\ldots,m$ must
have rank less than or equal to $n-1$. This condition can be
imposed by requiring \cite{horn:1990a} that all the $(n\times
n)$-submatrices of $R$ have zero determinant. If all the entries
of $R$ were polynomials\footnote{By which we mean polynomials in
the individual components of the vector $\mathbf{x}_j$.} in
$\mathbf{x}_1,\ldots,\mathbf{x}_m$ then this condition becomes the
requirement that a set of polynomials vanish.

Using the ingredients discussed in the previous paragraph we are
now able to express an optimal decomposition for a point
$\mathbf{r}\in\conv V$ as the simultaneous solution to a set of
polynomials. The key observation is that it is possible to take
the components of the basis vectors $\alpha_1,\ldots,\alpha_t$ for
the tangent space $TV$ of an algebraic variety $V$ (embedded in
affine space) to be polynomials in the coordinate vector
$\mathbf{x}\in V$. To see this recall \cite{harris:1995a} that the
tangent space to $V$, or the \emph{Zariski tangent space}, is
defined to be the null space of the matrix $J_{jk}\triangleq
\frac{\partial \ell_j}{\partial x_k}$, where $V =
\mathcal{Z}(\ell_1,\ldots,\ell_a)$. Because the entries of $J$ are
polynomials in the components of $\mathbf{x}$, the components of
the basis vectors $\alpha_j$ for the null space of $J$ can be
taken to be polynomials in $\mathbf{x}$ as well. (To see this,
find the basis for the null space of $J$ via row reduction. The
entries will be rational functions of $x_j$. To get polynomial
entries just clear denominators.)

We now return to the matrix $R$ which we introduced in the
previous two paragraphs. This matrix can be taken to have $m$ rows
consisting of the vectors\footnote{The other vectors
$\mathbf{x}_j-\mathbf{x}_k$, $j\not=(k+1)$ are linearly dependent
with this set.} $\mathbf{x}_j-\mathbf{x}_{j+1}$, $j=1,\ldots,m$,
where we identify $j=1$ with $j=m+1$, as well as $mt$ rows
consisting of the basis vectors
$\alpha_1^{(j)},\ldots,\alpha_t^{(j)}$, $j=1,\ldots, m$. There are
$\left(\begin{smallmatrix}m(t+1)\\
n\end{smallmatrix}\right)\left(\begin{smallmatrix}m\\
n\end{smallmatrix}\right)$ $n\times n$-submatrices of $R$. We
denote the determinants of each of these submatrices by
$\Delta_\alpha(\mathbf{x}_1,\ldots,\mathbf{x}_m)$, where $\alpha$
runs over all the $\left(\begin{smallmatrix}m(t+1)\\
n\end{smallmatrix}\right)\left(\begin{smallmatrix}m\\
n\end{smallmatrix}\right)$ possible submatrices. In order for a
hyperplane to be $m$-tangent to $\gr(f)$ we require that all of
the polynomials $\Delta_\alpha(\mathbf{x}_1,\ldots,\mathbf{x}_m)$
simultaneously vanish. By adjoining the equations
$\ell_j(\mathbf{x}_k)$, $j=1, \ldots, a$, $k=1, \ldots, m$ (which
ensure that $\mathbf{x}_j$ lie in $V$) and the equations
$\mathbf{0}=\sum_{j=1}^{m} p_j \mathbf{x}_j  - \mathbf{r}$,
$0=\sum_{j=1}^m p_j -1$ (which ensure that $\mathbf{r}$ lies in
the convex hull of $\mathbf{x}_j$) we obtain a set of conditions
for $\{p_j,\mathbf{x}_j\}$ to be an optimal $m$-decomposition for
$\mathbf{r}\in\conv V$. This discussion constitutes a proof of the
following proposition.

\begin{proposition}\label{thm:optdec}
An optimal $m$-decomposition for $\mathbf{r}\in\conv V$ consists
of $m$ points $\mathbf{x}_j\in\mathbb{R}^{n+1}$ and $m$ variables
$p_j$ which satisfy the following equations:
\begin{align}
0&=\ell_j(\mathbf{x}_k),\quad \forall j=1,\ldots,a, \forall
k=1,\ldots,m,\\
0&=\Delta_\alpha(\mathbf{x}_1,\ldots,\mathbf{x}_m), \quad\forall
\alpha = 1,\ldots,\left(\begin{smallmatrix}m(t+1)\\
n\end{smallmatrix}\right)\left(\begin{smallmatrix}m\\
n\end{smallmatrix}\right), \\
\mathbf{0}&=\sum_{j=1}^{m} p_j \mathbf{x}_j  - \mathbf{r},\\
0&=\sum_{j=1}^m p_j -1.
\end{align}
\end{proposition}

Proposition~\ref{thm:optdec} has two important consequences for
the evaluation of convex roofs. The first is that it provides an
\emph{implicit} representation for the optimal decomposition of a
point $\mathbf{r}\in\conv V$ for $\conv f$. This is in contrast to
standard procedure to calculate an optimal decomposition, i.e.,
via direct minimisation of (\ref{eq:caraconvroof}). In general
this minimisation must be performed numerically and, as a
consequence, there is no guarantee that a calculated decomposition
is optimal (It could just be a local minima). The polynomials of
Proposition~\ref{thm:optdec} therefore provide a
\emph{certificate} of optimality.

The second consequence of Proposition~\ref{thm:optdec} is that the
space required to search through to calculate an optimal
decomposition is dramatically reduced. Often, in practice, an
infinite search space is replaced with a finite search space.

To illustrate Proposition~\ref{thm:optdec} we write out the
equations that must be solved to find a bitangent plane for the
example introduced in Figure~\ref{fig:frame}. These equations read
\begin{align}
0&=x_1^2+y_1^2-1, \\
0&=x_2^2+y_2^2-1, \\
0&=\begin{vmatrix}
y_1 & -x_1 & x_1^2y_1 \\
y_2 & -x_2 & x_2^2y_2 \\
 x_1-x_2 & y_1-y_2 & x_1^3-x_2^3
\end{vmatrix},\\
0&=px_1 + (1-p)x_2-r_x,\\
0&=py_1 + (1-p)y_2-r_y.
\end{align}
By solving these equations we were able to calculate the convex
hull and convex roof illustrated in Figure~\ref{fig:convhull} and
Figure~\ref{fig:roof}.

\section{The Geometry of State Space: the Poincar\'e Sphere}\label{sec:poincare}

In this paper we are interested in calculating roofs for certain
scalar-valued functions defined on the state-space of quantum
systems. Initially, it seems that this problem is unrelated to the
results of the previous section because state-space (a Hilbert
space) has no obvious geometric character. In this section we show
that in fact pure state-space for quantum systems of dimension $D$
may be represented by the common zero-loci of a collection of
polynomial equations. This provides the first piece of information
--- the domain $V$ for the function $f$ we wish to roof --- needed to
apply the results of \S\ref{sec:prelims} to our problem. (We
discuss the second piece of information, the construction of the
function $f$ we wish to roof, in the next section.) We note that
there are many papers dealing with generalisations of the Bloch
sphere construction for higher-dimensional quantum systems. The
generalisation we recall here seems to date from 1997, where it
was considered in detail in the papers of \cite{khanna:1997a,
arvind:1997a}. The observation that this embedding of state-space
into $\mathbb{R}^n$ is a quadratic variety is, to the best of our
knowledge, original.

\subsection{Poincar\'e Spheres Based on $\mathfrak{su}(D)$}

The Bloch sphere representation for a rank-$2$ density
operator\footnote{Any rank-$2$ density operator
$\rho=p|v_1\rangle\langle v_1| + (1-p)|v_2\rangle\langle v_2|$,
where $|v_j\rangle$ are the eigenvectors of $\rho$, may be thought
of as the state of an effective two-level system (qubit) whose
states are ``logical zero'' $|0\rangle \triangleq |v_1\rangle$ and
``logical one'' $|1\rangle \triangleq |v_2\rangle$.} is a
fundamental conceptual tool in quantum mechanics (see, for
example, \cite{nielsen:2000a} and \cite{preskillnotes} for a
discussion of the Bloch sphere). The fact that the space of all
rank-$2$ density operators corresponds to a geometrical object as
simple as the sphere $S^2$ and its interior in $\mathbb{R}^3$ has
led to many discoveries. This is due, in part, to the fact that it
is possible to employ geometric intuition when dealing with
systems whose states or dynamics can be represented on the Bloch
sphere. It is therefore very desirable to consider generalisations
of the Bloch sphere to see if there is a simple geometrical object
that represents the state-space for a $D$-level system.
Unfortunately there is no object which possesses quite the
simplicity of the Bloch sphere. However, as we argue, there is one
natural way to generalise the Bloch sphere construction which is
useful when discussing certain aspects of rank-$D$ density
operators.

In this section we discuss a generalisation of the Bloch sphere
representation for density operators of rank-$D$, which we call
the Poincar\'e sphere (we choose this terminology in order to
avoid confusion with the Bloch sphere which specifically refers to
$S^2$). The first step in the Poincar\'e sphere construction is to
choose an operator basis with respect to which an arbitrary
rank-$D$ density operator can be expanded. We choose the set of
traceless hermitian generators\footnote{These matrices are also
known as the \emph{Gellmann matrices}.} $\lambda^j$,
$j=1,\ldots,D^2-1$ for the group $\text{\emph{SU}}(D)$. (As we
will show, this choice is entirely arbitrary. Any other basis for
the operator space may be obtained via a linear transformation
from this one.) We represent the generators in an orthonormal
basis $\{|v_a\rangle\}$.  The generators are given in terms of the
set of operators $\boldsymbol{\Gamma}\triangleq\left\{\Gamma_a,
\Gamma^{(+)}_{ab}, \Gamma^{(-)}_{ab}\right\}$ defined by
\begin{equation}
\label{eq:sulam1} \Gamma_a \triangleq \frac{1}{\sqrt{a(a-1)}}
\Biggl(\sum_{b=1}^{a-1}|v_b\rangle\langle v_b| -
(a-1)|v_a\rangle\langle v_a|\Biggr), \quad 2\le a\le D,
\end{equation}

\begin{equation}
\label{eq:sulam2} \Gamma_{ab}^{(+)} \triangleq \frac{1}{\sqrt{2}}
(|v_a\rangle\langle v_b| + |v_b\rangle\langle v_a|),\quad 1\le
a<b\le D,\\
\end{equation}

\begin{equation}
\label{eq:sulam3} \Gamma_{ab}^{(-)} \triangleq -\frac{i}{\sqrt{2}}
(|v_a\rangle\langle v_b| - |v_b\rangle\langle v_a|),\quad 1\le
a<b\le D.
\end{equation}
The set $\boldsymbol{\Gamma}$ consists of $D^2-1$ hermitian
operators. We assume that the set $\boldsymbol{\Gamma}$ is
ordered, and we set $\lambda^j$ to be equal to the $j$th element
of $\boldsymbol{\Gamma}$.  Note that, in this paper, we suppose
that roman indices $j$ run from $1$ to $D^2-1$.

The hermitian operators $\lambda^j$ form a representation for the
Lie algebra $\mathfrak{su}(D)$, which allows us to
write\footnote{We are employing the Einstein summation convention
over repeated indices, so that ${d^{jk}}_{l} \lambda^l$ means
$\sum_{l=1}^{D^2-1}{d^{jkl}} \lambda^l$.}
\begin{equation}
\lambda^j\lambda^k = \frac1D\delta^{jk} + {d^{jk}}_{l} \lambda^l +
i{f^{jk}}_{l} \lambda^l,
\end{equation}
where ${f^{jk}}_{l}$ are the completely antisymmetric structure
constants of $\mathfrak{su}(D)$, and the completely symmetric
coefficients ${d^{jk}}_{l}$ are given by
\begin{equation}\label{eq:acomm}
\{\lambda^j,\lambda^k\} = \frac2D\delta^{jk} + 2{d^{jk}}_{l}
\lambda^l.
\end{equation}
By adjoining the operator
\begin{equation}
\lambda^0 = \frac{1}{\sqrt{D}}I
\end{equation}
to the $D^2-1$ hermitian generators we obtain a complete basis for
the space of all $D\times D$ hermitian operators.  Further, this
basis is orthonormal under the Hilbert-Schmidt inner product:
\begin{equation}
\tr(\lambda^\alpha\lambda^\beta)=\delta^{\alpha\beta}.
\end{equation}
(In the following greek indices are assumed to run from $0$ to
$D^2-1$.)

Using the operator basis (\ref{eq:sulam1})-(\ref{eq:sulam3}) we
can obtain a geometric representation for the space of all pure
states of a $D$-dimensional quantum system. Recall
\cite{peres:1993a} that a hermitian operator $\rho$ represents a
state (either pure or mixed) of a quantum system if it satisfies
the following conditions:
\begin{equation}\label{eq:psalgcond}
\tr(\rho)=1, \quad \rho^\dag=\rho, \quad \text{and} \quad
\rho^2\le\rho,
\end{equation}
where by $\rho^2\le\rho$ we mean that $\rho-\rho^2$ is positive
semidefinite and we require equality when $\rho$ is a pure state.
(These conditions assert that $\rho$ is positive semidefinite with
trace unity.) Utilising the fact that any density operator may be
expanded uniquely in terms of the operator basis as
\begin{equation}
\rho = \frac1D c_\alpha \lambda^\alpha,
\end{equation}
where the coefficients $c_\alpha\in\mathbb{R}$ are given by
\begin{equation}
c_\alpha = D\tr(\rho\lambda^\alpha),
\end{equation}
we can translate the pure-state conditions (\ref{eq:psalgcond})
into geometric conditions on $c_\alpha$.

Firstly, normalisation implies that
\begin{equation}
\rho = \frac{I+c_j\lambda^j}{D}.
\end{equation}
In this way the components $\mathbf{c}$ can be regarded as a
vector in a $(D^2-1)$-dimensional real vector space (this is where
we impose the hermiticity requirement), which we can identify with
$\mathbb{R}^{D^2-1}$. If the density operator is a pure state
$\rho = |\psi\rangle\langle\psi|$, then $\tr(\rho^2)=1$ implies
that
\begin{equation}\label{eq:pscond}
|\mathbf{c}|^2 = D(D-1).
\end{equation}

Before we complete our description of the geometric conditions we
need to introduce three vector operations on the vectors
$\mathbf{c}$ defining pure states. The first such operation is
simply the euclidean inner product,
\begin{equation}\label{eq:vinner}
\mathbf{a}\cdot\mathbf{b} = a_jb^j.
\end{equation}
The second operation is a generalisation of the cross-product
formula in $3$ dimensions, the \emph{wedge} or \emph{outer
product}, which is constructed using the antisymmetric structure
constants
\begin{equation}\label{eq:vouter}
(\mathbf{a}\wedge\mathbf{b})_l = a_jb_k{f^{jk}}_l.
\end{equation}
The third operation is a completely symmetric vector product,
\emph{the star product}
\begin{equation}\label{eq:vsymm}
(\mathbf{a}\star\mathbf{b})_l = a_jb_k{d^{jk}}_l.
\end{equation}
The vector operations (\ref{eq:vinner})-(\ref{eq:vsymm}) satisfy a
number of useful relations.  To see this, let
$U\in\text{\emph{SU}}(D)$ be a group element.  Let $O(U)$ be the
operator transforming in the adjoint representation of
$\text{\emph{SU}}(D)$ that represents $U$ whose matrix elements
are given by the formula
\begin{equation}
[O(U)]_{\alpha\beta} = \tr(\lambda^\alpha U \lambda^\beta U^\dag).
\end{equation}

The vector operations defined above transform in the following way
under the adjoint representation of $\text{\emph{SU}}(D)$,
\begin{align}
O(U)\mathbf{a}\cdot O(U)\mathbf{b} &= \mathbf{a}\cdot\mathbf{b}, \\
O(U)\mathbf{a}\wedge O(U)\mathbf{b} &=
O(U)(\mathbf{a}\wedge\mathbf{b}), \\
O(U)\mathbf{a}\star O(U)\mathbf{b} &=
O(U)(\mathbf{a}\star\mathbf{b}).
\end{align}
By definition the inner product is an
$\text{\emph{SO}}(D^2-1)$-scalar.  Both the star and wedge product
of two vectors transforming in the adjoint representation of
$\text{\emph{SU}}(D)$ give rise to another vector transforming in
the adjoint representation.

We can now state a geometric version of the final pure-state
condition. Equation (\ref{eq:pscond}) is not a sufficient
condition for the vector $\mathbf{c}$ to represent a pure state.
This is a consequence of the fact that the identity
$\tr(\rho^2)=1$ does not imply $\rho^2=\rho$ when $D>2$. The
further constraints due to the anticommutation relations
(\ref{eq:acomm}) must be taken into account. If these constraints
are taken into account then the following condition, along with
(\ref{eq:pscond}), is necessary and sufficient for the state
represented by $\mathbf{c}$ to be pure. The additional constraint
on $\mathbf{c}$ is known as the \emph{star-product} formula
\cite{khanna:1997a, arvind:1997a},
\begin{equation}
\mathbf{c} \star \mathbf{c} = (D-2)\mathbf{c}.
\end{equation}
With these constraints, the space of all pure states is the
subvariety $\mathcal{P}^\star\subset S^{D^2-1}$ which is the
common locus of zeros of the following set of quadratic
polynomials,
\begin{align}\label{eq:vanfun}
\ell_0(\mathbf{c}) &= |\mathbf{c}|^2-D(D-1),\\
\ell_l(\mathbf{c}) &= \mathbf{c}^T D^l \mathbf{c} - \mathbf{e}_l^T
\mathbf{c},\label{eq:psexcon}
\end{align}
where $D^l$ are real symmetric matrices with entries given by
$[D^l]_{jk} = {d^{jk}}_l$, and $\mathbf{e}_j$ is a row-vector
whose $j$th entry is one and the rest zero.

The space $\mathcal{P}$ of all pure states may be envisioned as a
section $\mathcal{P}^\star$ of the sphere $S^{D^2-1}$. The space
of all mixed states $\mathcal{D}^\star$ in this representation
(\emph{the Poincar\'e sphere}) is simply given by the convex hull
of this quadratic variety, $\mathcal{D}^\star = \conv
\mathcal{P}^\star$.

To get an idea of what $\mathcal{P}^\star$ looks like as a
subspace of $S^{D^2-1}$ consider the expression
\begin{equation}
\tr(\rho\sigma) = \frac{1}{D} +
\frac{\mathbf{r}\cdot\mathbf{s}}{D^2},
\end{equation}
where $\rho = \frac{I+\mathbf{r}\cdot\boldsymbol{\lambda}}{D}$ and
$\sigma = \frac{I+\mathbf{s}\cdot\boldsymbol{\lambda}}{D}$ are any
two pure states. Using the formula $\mathbf{r}\cdot\mathbf{s} =
|\mathbf{r}||\mathbf{s}|\cos{\theta}$ we find that the angle
$\theta$ between the vectors representing $\rho$ and $\sigma$ is
given by
\begin{equation}
\theta = \cos^{-1}\left(\frac{D\tr(\rho\sigma)-1}{D-1}\right).
\end{equation}
This angle is largest when $\tr(\rho\sigma)$ is zero.  In this way
we see that widest angle $\theta_{\max}$ between any two points in
$\mathcal{P}^\star$ is given by
$\cos^{-1}\left(\frac{1}{1-D}\right)$, so that if $\mathbf{r}$
represents a pure quantum state, $-\mathbf{r}$ cannot, unless
$D=2$.

\subsection{Poincar\'e Spheres from General Operator Bases}

In the previous subsection we introduced a geometric
representation for the convex set of all density operators for a
$D$ dimensional quantum system. In order to do so we expanded an
arbitrary density operator $\rho$ in terms of a specific operator
basis, the generalised Gellmann matrices. In this subsection we
briefly summarise the analogous results for when we expand $\rho$
in terms of an arbitrary (hermitian) operator basis.

Our main motivation for investigating Poincar\'e sphere
representations for $\rho$ with respect to different operator
bases is to better study states $\rho$ of a multipartite system.
Because we are interested in situations when a pure-state density
operator $\rho$ is a product state $\rho=\sigma\otimes\omega$, it
is desirable, in some cases, to expand $\rho$ in terms of an
operator basis which makes this property manifest, such as the
basis $\{\lambda^\alpha\otimes\lambda^\beta\}$ formed from the
tensor product of two Gellmann bases.

If we expand a density operator $\rho$ in terms of an operator
basis different from the Gellmann matrices we find that the
generalised Poincar\'e sphere so derived is different. It turns
out however, that the new Poincar\'e sphere is related to the
original sphere via an overall isometry. To see this we need to
derive a more general geometric form of the pure state conditions
(\ref{eq:psalgcond}). For clarity we limit ourselves to complete
hermitian orthonormal operator bases $\mu^\alpha$, $\alpha =
1,\ldots, D^2$.

The geometric pure state conditions analogous to (\ref{eq:vanfun})
and  (\ref{eq:psexcon}) follow from expanding the (hermitian)
product $\mu^\alpha\mu^\beta$ in terms of the complete basis,
\begin{equation}
\mu^\alpha\mu^\beta = {\chi^{\alpha\beta}}_\gamma \mu^\gamma,
\end{equation}
where ${\chi^{\alpha\beta}}_\gamma$ are the expansion
coefficients. Following the analysis detailed in the previous
subsection we can write out the pure state conditions in terms of
$c_\alpha$:
\begin{align}
c_\alpha \tau^\alpha = 1, \label{eq:psgencon1}\\
c_\alpha c^\alpha = 1, \label{eq:psgencon2}\\
c_\alpha c_\beta {\chi^{\alpha\beta}}_\gamma = c_\gamma,
\label{eq:psgencon3}
\end{align}
where $\tau^\alpha = \tr(\mu^\alpha)$. The pure state conditions
(\ref{eq:psgencon1}), (\ref{eq:psgencon2}), and
(\ref{eq:psgencon3}) say that $c_\alpha$ must be a vector in
$\mathbb{R}^{D^2}$ which lies in the the intersection of the
hypersphere (\ref{eq:psgencon2}) with a hyperplane
(\ref{eq:psgencon1}) and $D^2$ quadric surfaces
(\ref{eq:psgencon3}).

The algebraic variety $V$ defined by (\ref{eq:psgencon1}),
(\ref{eq:psgencon2}), and (\ref{eq:psgencon3}) representing the
space of all pure states of a $D$-dimensional quantum system is
related to the $\mathfrak{su}(D)$ Poincar\'e sphere in a simple
way. Any complete orthonormal hermitian operator basis
$\mu^\alpha$ can be found as an orthogonal rotation of the
$\mathfrak{su}(D)$ basis,
\begin{equation}
\mu^\alpha = {O^\alpha}_\beta \lambda^\beta,
\end{equation}
where ${O^\alpha}_\beta\in\text{\emph{SO}$(D^2)$}$. Say we wish to
represent $\rho$ with respect to some other operator basis, i.e.\
write $\rho = c_\alpha\mu^\alpha = d_\alpha \lambda^\alpha$, and
we want to know how the geometric pure state conditions for
$c_\alpha$ and $d_\alpha$ are related. This relationship can be
elucidated by making use of the formula $c_\alpha = d_\alpha
{O^\alpha}_\beta$. By substituting this formula into the geometric
pure state conditions (\ref{eq:psgencon1}), (\ref{eq:psgencon2}),
and (\ref{eq:psgencon3}) we find that $d_\alpha$ must satisfy
\begin{align}
d_\alpha {\tau'}^\alpha = 1, \label{eq:psgemcon1}\\
d_\alpha d^\alpha = 1, \label{eq:psgemcon2}\\
d_\alpha d_\beta {{\chi'}^{\alpha\beta}}_\gamma = d_\gamma,
\label{eq:psgemcon3}
\end{align}
where ${\tau'}^\alpha = {O^\alpha}_\beta\tau^\beta$ and
${{\chi'}^{\alpha\beta}}_\gamma =
{O^\alpha}_{\alpha'}{O^\beta}_{\beta'}{{\chi'}^{{\alpha'}{\beta'}}}_{\gamma'}{O^
{\gamma'}}_{\gamma}$.  (When $\lambda^\alpha$ is the Gellmann
basis these conditions reduce to (\ref{eq:vanfun}) and
(\ref{eq:psexcon}).) In this way we see that the generalised
Poincar\'e representation is unique up to an isomorphism of
varieties. The geometric representation of the space of all
rank-$D$ mixed states is simply given by the convex hull of the
variety $V$ defined by (\ref{eq:psgemcon1}), (\ref{eq:psgemcon2}),
and (\ref{eq:psgemcon3}). We note, in passing, that the variety
$V'$ defined by (\ref{eq:psgemcon3}) is a geometric representation
of the extreme points of the positive cone of nonnegative $D\times
D$ matrices. The space of all positive linear combinations of
points in $V'$ is a geometric representation for the positive cone
of nonnegative $D\times D$ matrices.

\section{Entanglement Measures}\label{sec:entmeas}

In this section we briefly review the definitions and some results
that have been obtained in the theory of entanglement. In
particular we introduce a set of axioms for a function to be
considered an entanglement measure. We then introduce a family of
polynomial functions which satisfy these axioms. Finally, we
demonstrate that the method of \S\ref{sec:prelims} can be applied
to evaluate these functions.

Suppose we have a bipartite quantum system $AB$ composed of two
subsystems $A$ and $B$ whose respective Hilbert spaces,
$\mathcal{H}_A$ and $\mathcal{H}_B$, have dimensions
$\dim\mathcal{H}_A = d_A$ and $\dim\mathcal{H}_B = d_B$. Our
problem, at its most elementary level, is to answer questions such
as: given a pure state $\rho=|\psi\rangle\langle\psi|$ of $AB$,
when is this state expressible as a tensor product
$\rho=\sigma\otimes\omega$? More precisely, one of the general
problems we are interested in is the following. Given an arbitrary
mixed state $\rho$ of a multipartite quantum system $A_1A_2\cdots
A_n$, when is $\rho$ expressible as a convex combination of
tensor-product states:
\begin{equation}\label{eq:mpartysep}
\rho \overset{?}{=} \sum_{j=1}^{s\ge \rank\rho} p_{j}
\sigma_{A_1}^{(j)}\otimes
\sigma_{A_2}^{(j)}\otimes\cdots\otimes\sigma_{A_n}^{(j)},
\end{equation}
where $\sigma_{A_k}^{(k)}$ are pure states? This problem is
fundamental in the theory of entanglement and has received a great
deal of attention in recent times. This work has culminated in the
development of several criteria to determine when a state can be
written in the form (\ref{eq:mpartysep}).

Deciding if a mixed state $\rho$ is expressible as in
(\ref{eq:mpartysep}) is significantly complicated by the fact that
there are many ways to write a mixed state as a convex combination
of pure states. Fortunately there is a recipe which can be
utilised to relate all the different decompositions of $\rho$ in
terms of convex mixtures of pure states \cite{schrodinger:1936a,
jaynes:1957a, hughston:1993a}. To describe this recipe we first
write $\rho$ in terms of the decomposition
\begin{equation}\label{eq:eofmin}
\rho = \sum_{j=1}^{r} p_j |u_j\rangle\langle u_j| = \sum_{j=1}^{r}
|\overline{u}_j\rangle\langle \overline{u}_j|.
\end{equation}
where $r$ is the rank of $\rho$, $\{|u_j\rangle\}$ are the
eigenvectors of $\rho$ and $|\overline{u}_j\rangle =
\sqrt{p_j}|u_j\rangle$ are the subnormalised eigenvectors of
$\rho$. It turns out that any other decomposition of $\rho$ can be
found via the equation
\begin{equation}\label{eq:eofmin2}
\rho = \sum_{j=1}^{s\ge r} |\overline{v}_j\rangle\langle
\overline{v}_j|,
\end{equation}
where $|\overline{v}_j\rangle = \sum_{k=1}^{r}U^*_{jk}
|\overline{u}_k\rangle$ and $U$ is any right-unitary matrix. (The
number $s$ in Eq.~(\ref{eq:eofmin2}) may take any integer value
greater than or equal to the rank of $\rho$.  This is because
$\rho$ may be realised from an ensemble consisting of arbitrarily
many pure states.)

The problem described in the previous paragraphs is known as the
\emph{separability problem} --- a state $\rho$ being
\emph{separable} if it is expressible as in (\ref{eq:mpartysep}),
and \emph{entangled} otherwise. By itself this is an interesting
mathematical problem. However, there are deeper reasons for why we
are interested in answering it. These stem from the fact that
\emph{entanglement}, the property possessed by entangled states,
seems to be a fundamentally new kind of resource like free energy
or information. Motivated by this, researchers are interested in
working out when a state contains some of this resource (i.e.\
deciding when $\rho$ is not separable). Ultimately, however,
researchers are more interested in working out \emph{how much} of
this resource is present in a given state.

In this area of research, the theory of entanglement, a unified
approach to answering questions such as the separability of a
state $\rho$ and the measurement of the entanglement of $\rho$ has
been developed. This approach is based on studying the properties
of certain functions, called \emph{entanglement measures}, on
state-space which are zero for all separable states, and which are
nonzero for all other states.

At the current time there are only hints that entanglement
measures actually measure the resource character of
entanglement\footnote{This has only been conclusively established
for the entanglement for pure-states of a bipartite system
\cite{bennett:1996a, popescu:1997a, vidal:2000b, nielsen:2000b}.}.
We ignore this problem for the time being, concentrating on the
more basic problem of evaluating certain entanglement measures.

\subsection{Pure-State Entanglement Measures}

What properties must a scalar-valued function
$F:\mathcal{H}\rightarrow \mathbb{R}$ possess in order for it to
be described as an entanglement measure? This problem has been
widely studied in the physics literature and a partial consensus
has been reached\footnote{To date, the physics behind the
measurement of entanglement for multipartite quantum systems has
not been satisfactorily elucidated. Only in the bipartite case has
agreement been reached. The conditions we use for a function to be
a multipartite entanglement measure follow \cite{vidal:2000b}.}.

For bipartite quantum systems with state-space isomorphic to
$\mathbb{C}^M\otimes\mathbb{C}^N$, with $N\le M$, it has been
shown \cite{vidal:2000b} that entanglement measures are in
bijective correspondence with unitarily invariant, concave (and
hence continuous) real-valued functions on the space of density
operators over $\mathbb{C}^M$. More precisely, consider any
(continuous) real-valued function
$f:\mathcal{D}(\mathbb{C}^M)\rightarrow\mathbb{R}$ on the space
$\mathcal{D}(\mathbb{C}^M)$ of density operators $\rho$ on
$\mathbb{C}^M$ which satisfies:
\begin{enumerate}
\item invariance under the transformation of $\rho$ by any unitary
matrix $U:\mathbb{C}^M\rightarrow\mathbb{C}^M$, i.e.\
\begin{equation}
f(U\rho U^\dag) = f(\rho), \quad \text{and}
\end{equation} %
\item concavity, i.e.\
\begin{equation}
f(\rho) \ge pf(\omega_1) + (1-p)f(\omega_2),
\end{equation}
where $\rho = p\omega_1 + (1-p)\omega_2$, $p\in[0,1]$, and
$\omega_1,\omega_2\in\mathcal{D}(\mathbb{C}^M)$.
\end{enumerate}
Using $f$ a (pure-state) entanglement measure $F$ on $\mathcal{H}$
can be constructed by defining
\begin{equation}
F(|\phi\rangle)\triangleq f(\tr_B(|\phi\rangle\langle\phi|)),
\end{equation}
where $\tr_B$ denotes the partial trace over subsystem $B$.

Using the correspondence between pure-state bipartite entanglement
measures and unitarily invariant concave functions we propose a
family of entanglement measures using for $f$ the functions
\begin{equation}
f_a(\rho) \triangleq 2(1-\tr(\rho^a)),
\end{equation}
where $a\in\mathbb{N}, a>1$. The functions $f_a$ are easily seen
to be unitarily invariant and concave. The family $f_a$ has one
further property, namely that they are zero for pure states
$\rho$. This means that the entanglement measures $F_a$ that they
give rise to under the construction discussed above are zero if
and only if $|\phi\rangle$ is a product state.

If we use the Poincar\'e sphere representation of $\mathcal{H}$ we
see that $f_\alpha$ are polynomial functions of the coordinate
vector $c_\alpha$ for a state $|\phi\rangle$. This can be
established as follows. First expand the state $|\phi\rangle$ in
terms of the complete orthonormal operator basis given by
$\{\lambda^\alpha\otimes \lambda^\beta\,|\,\alpha = 1,\ldots,M^2,
\beta = 1,\ldots,N^2 \}$, which is the tensor product of the
Gellmann matrices of sides $M$ and $N$ respectively:
\begin{equation}
\rho = c_{\alpha\beta}\lambda^\alpha\otimes \lambda^\beta.
\end{equation}
Taking the partial trace over $B$ yields
\begin{equation}
\tr_B(\rho) = c_{\alpha 0} \lambda^\alpha,
\end{equation}
because $\lambda^{\alpha}$ is traceless for $\alpha\not=0$. If we
now calculate $f_a$ we find that
\begin{equation}\label{eq:altformpe}
f_a(|\phi\rangle) = 2(1-\tr( (c_{\alpha 0} \lambda^\alpha)^a )),
\end{equation}
which is a polynomial in the coefficients $c_{\alpha\beta}$. This
function is defined on the Poincar\'e sphere for the
tensor-product operator basis $\{\lambda^\alpha\otimes
\lambda^\beta\}$. The entanglement measure $f_a$ can also be
written as a polynomial in terms of the position vector $d_\alpha$
defining $\rho$ in any other operator basis by replacing
$c_{\alpha0}$ with ${O_{\alpha0}}^{\beta}d_\beta$, where $O$ is
the $\text{\emph{SO}}(M^2N^2)$ operator connecting the two
operator bases.

\subsection{Mixed-State Entanglement Measures from the Roof Construction}
In the previous subsection we discussed a set of conditions for a
function on state-space to be considered an entanglement measure.
We introduced a family of entanglement measures which are
polynomial functions of the vector representing a state in the
Poincar\'e sphere representation. In this subsection we discuss a
technique to construct entanglement measures defined for all
states, both pure and mixed, from pure-state entanglement
measures.

Any mixed state $\rho$ can be written as a convex combination
$\{p_j,|\phi_j\rangle\}$ of pure states,
\begin{equation}
\rho = \sum_{j=1}^{r} p_j |\phi_j\rangle\langle\phi_j|.
\end{equation}
As we discussed at the beginning of this section this
representation is not unique. In order to measure the entanglement
of a mixed state $\rho$ it is necessary to take account of this
nonuniqueness of the pure state decomposition. Each decomposition
of $\rho$ in terms of pure states represents a way of
manufacturing the mixed state $\rho$, because
$\rho=\sum_{j=1}^{r}p_j|\phi_j\rangle\langle\phi_j|$ can be
thought of as being produced from a quantum source emitting the
pure state $|\phi_j\rangle$ with probability $p_j$. If we now
imagine that entanglement is an expensive quantity that ``costs''
$F(|\phi_j\rangle)=f(\tr_B(|\phi_j\rangle\langle\phi_j|))$, then
it is natural to ask what is the minimal cost of producing $\rho$.
This leads us to define a quantity which is called the
(generalised) \emph{entanglement of formation} (see
\cite{bennett:1996a} for the original definition and further
discussions):
\begin{equation}
\mathcal{F}(\rho) \triangleq \inf_{\{q_j,|\psi_j\rangle\}}
\sum_{j=1}^{r\ge s\ge r^2} q_j
f(\tr_B(|\psi_j\rangle\langle\psi_j|)),
\end{equation}
where $\rho=\sum_{j=1}^s q_j|\psi_j\rangle\langle\psi_j|$, which
quantifies the minimum expected cost\footnote{It should be noted
that the interpretation of $\mathcal{F}$ as quantifying the cost
of producing the state $\rho$ is subject to the condition that the
entanglement of formation is additive on tensor products.} of
preparing $\rho$.

If we translate the definition of the mixed state entanglement
measure $\mathcal{F}$ into the Poincar\'e sphere representation we
find that it coincides precisely with taking a roof of the scalar
valued function $f$. In this way the results of
\S\ref{sec:prelims} can be applied to calculate $\mathcal{F}=\conv
f$.

\section{Conclusions}
In this paper we have considered the problem of calculating the
convex hull of certain class of affine algebraic varieties. We
have shown that this problem is intimately related to a problem
studied in quantum information science, namely that of calculating
mixed-state entanglement measures. Our principle result is that an
optimal decomposition for a point lying in the roof of a function
$f$ must satisfy a set of polynomial equations. This provides an
\emph{implicit} representation for such decompositions.

It is tempting to apply the well-developed techniques in
computational commutative algebra to calculate the polynomial
entanglement measures we introduced in \S\ref{sec:entmeas} for
mixed states. Unfortunately, apart from the example we considered
in Figures~\ref{fig:frame}, \ref{fig:convhull}, and \ref{fig:roof}
(which pertains to a cubic entanglement measure for \emph{real}
rank-$2$ density operators) the polynomials that appear in
Proposition~\ref{thm:optdec} are far too complex for current
computational approaches.

One principle future direction suggests itself at this stage.
Determining the \emph{additivity} of convex functionals
(particularly entanglement of formation-like quantities) on tensor
products of quantum states is a major unsolved problem in quantum
information science. It is plausible that the implicit
representation afforded by Proposition~\ref{thm:optdec} may
provide one route to considering this problem.

\subsection*{Acknowledgements}
I would like to thank Michael Nielsen and Bill Wootters for many
inspiring conversations which led to this work. I would also like
to thank Armin Uhlmann for many helpful suggestions and for
communicating some unpublished results on roofs which allowed me
to fix some problems in a previous version of this paper. I am
also grateful to the EU for support for this research under the
IST project RESQ

\newcommand{\etalchar}[1]{$^{#1}$}
\providecommand{\bysame}{\leavevmode\hbox
to3em{\hrulefill}\thinspace}

\end{document}